\documentclass[twocolumn]{emulateapj}
\usepackage{subfigure}
\begin{document}
\title{Compositions of Hot Super-Earth Atmospheres: exploring Kepler Candidates}
\author{Y. Miguel}
\affil{Max Planck Institut fuer Astronomie, Koenigstuhl 17, 69117, Heidelberg, Germany}
\email{miguel@mpia-hd.mpg.de}
\author{L. Kaltenegger} 
\affil{Max Planck Institut fuer Astronomie, Koenigstuhl 17, 69117, Heidelberg, Germany\\
Harvard Smithsonian Center for Astrophysics, 60 Garden St., 02138 MA,Cambridge, USA}
\author{B. Fegley Jr.} 
\affil{Planetary Chemistry Laboratory, Dpt. of Earth \& Planetary Sciences and Mc Donnell Center for the Space Sciences, Washington University, St. Louis, MO 63130, USA}
\author{L. Schaefer} 
\affil{Harvard Smithsonian Center for Astrophysics, 60 Garden St., 02138 MA,Cambridge, USA}

\begin{abstract}

This paper outlines a simple approach to evaluate the atmospheric composition of hot rocky planets by assuming different types of planetary composition and using corresponding model calculations. To explore hot atmospheres above 1000 K, we model the vaporization of silicate magma and estimate the range of atmospheric compositions according to the planet's radius and semi-major axis for the {\it{Kepler}} February 2011 data release. Our results show  5 atmospheric types for hot, rocky super-Earth atmospheres, strongly dependent on the initial composition and the planet's distance to the star. We provide a simple set of parameters that can be used to evaluate atmospheric compositions for current and future candidates provided by the {\it{Kepler}} mission and other searches.

\end{abstract}
\section{Introduction}

In recent years the search for extrasolar planets has resulted in the discovery of the first confirmed rocky planets with minimum masses below 10 M$_{\oplus}$ (hereafter super-Earths) \citep{ud07,le09} and radii below 2 R$_{\oplus}$, consistent with rocky planet models \citep{bo11,ba11}. Many of the detected super-Earths are located extremely close to their stars, at distances much less than Mercury to the Sun. These close-in hot super-Earths can reach temperatures high enough to melt the silicates present on their surfaces by irradiation from their host star alone. This causes outgassing and the subsequent formation of the planet's atmosphere. 

The formation of a rocky planet's atmosphere is a complex process built up from initial capture and degassing of material of meteoritic composition. Our model is based on work by \citet{scf04, scf09} and assumes, as a first order approximation, that at temperatures $T_p>$1000 K, the material present on the surface of the planet vaporizes and forms an outgassed atmosphere. In this process, the atmosphere is strongly dependent on the planetary composition \citep{scf09,va10}. \citet{scf09} modeled the specific case of CoRoT-7b and showed that these close-in planets may have a silicate atmosphere as a result of outgassing of magma on the surface. 

 The {\it{Kepler}} mission recently announced 1235 planetary candidates \citep{bo11}, of which 615 are consistent with models of rocky interior ($R<2 \pm$ 0.5 R$_{\oplus}$, accounting for the 25\% error in radii in the Kepler data set).  193 of these {\it{Kepler}} objects of interest (KOI) have potentially extremely high temperatures on their surfaces ($T_p>$1000 K, when assuming a 0.01 albedo), due to the intense stellar irradiation to which they are exposed, because of their proximity to the central star. 

In this work, we explore the composition of initial planetary atmospheres of hot super-Earths in the Kepler planet candidate sample, according to their semi major axis. Our approach can be applied to current and future candidates provided by the Kepler mission and other exoplanet searches.

\section{Outgassing Model}

We use the MAGMA code \citep{fc87,scf04} which calculates the equilibrium between the melt and vapor in a magma exposed at temperatures higher than 1000 K \citep{ha82,ha84,hb85,hb86}, for Al, Ca, Fe, K, Mg, Na, O, Si, Ti and their compounds. 

 We assume that the initial composition of a hot super-Earth is similar to the composition of komatiites, which are ultramafic lavas that were erupted during the Archaean on Earth (3.8 to 2.5 billion years ago), when Earth had a higher surface temperature. This choice is also supported by the work by \citet{rgf04}, who showed, for the case of the Earth and Mars, that a more massive planet has a lower composition of FeO in the mantle than a less massive one, a consequence of the solubility of oxygen in liquid iron-alloy,  which increases with temperature.  Table \ref{composicion} shows komatiite and bulk silicate Earth (BSE) compositions (see section \ref{section:composiciones}) for comparison. 

\begin{deluxetable*}{lccccccccr} 
\tablecolumns{10} 
\tablewidth{0pc} 
\tablecaption{Compositions adopted for the magma} 
\tablehead{
\colhead{Oxide} & \colhead{SiO$_2$} &\colhead{MgO} &\colhead{Al$_2$O$_3$} &\colhead{TiO$_2$} &\colhead{Fe$_2$O$_3$} &\colhead{FeO} &\colhead{CaO} &\colhead{Na$_2$O} &\colhead{K$_2$O} }
\startdata
Komatiite (\%) & 47.10 & 29.60 &  4.04 &  0.24 & 12.80 &  0.0 &  5.44 &  0.46 & 0.09 \\
Bulk Silicate Earth (\%) & 45.97 & 36.66 & 4.77 & 0.18 & 0.0 & 8.24 & 3.78 & 0.35 & 0.04\\
\enddata 
\label{composicion}
\end{deluxetable*} 

\section{Atmospheric Composition of Kepler Candidates}\label{kepler}

The planet's equilibrium temperature due to irradiation by a central star of radius $R_{\star}$ and effective temperature $T_{eff,\star}$, is given by equation \ref{tp},

\begin{equation}\label{tp}
 T_p=T_{eff,\star}\bigg(\frac{(1-A)R_{\star}^2}{4a^2}\bigg)^{\frac{1}{4}}
\end{equation}
where $a$ is the planet's semi-major axis and $A$ its bond albedo. We adopt an albedo of 0.01 \citep{mi10}.  Different values for the albedo can change our results slightly (see discussion). Here we set $T_p$ equal to the planet's initial surface temperature, to explore the resulting atmospheric composition.

\begin{figure}
\includegraphics[angle=270,scale=.33]{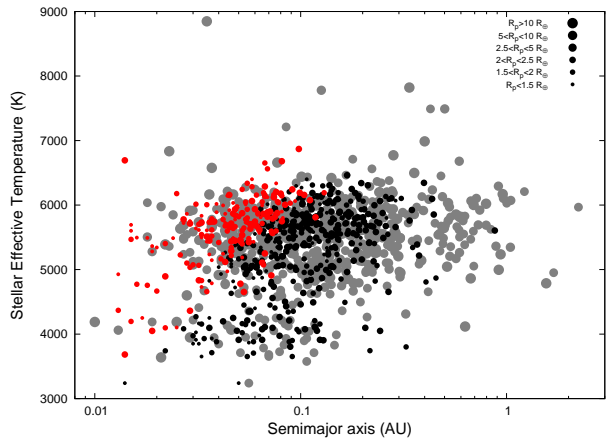}
\caption{Effective stellar temperature vs semi-major axis of all {\it{Kepler}} planetary candidates, in different point sizes, according to their radii. Planet candidates with $T_p>$1000 K and $R_p \le$ 2.5 R$_{\oplus}$, are shown in red. The cooler planets (in the same radius range) are shown in black. The planet candidates with $R_p>$2.5 R$_{\oplus}$ are shown in grey.}
\label{fig1}
\end{figure}

Figure \ref{fig1} shows the stellar effective temperature versus planet semi-major axis for all the {\it{Kepler}} candidates, where different point sizes indicate different planet radius. Using equation \ref{tp}, we separate the planets according to their radius and surface temperature. The planet candidates with derived $T_p>$ 1000 K and $R_p \le$ 2.5 R$_{\oplus}$, are shown in red and are the sample we focus on here. The cooler planets (in the same radius range) are shown in black. The planet candidates with $R_p>$2.5 R$_{\oplus}$ are shown in grey.

\subsection{Simulation Results}

We perform outgassing simulations for temperatures between 1000 and 3500 K, and apply our results to planet candidates in the February 2011 {\it{Kepler}} release with $R_p\le$ 2.5 R$_{\oplus}$ and $T_p>1000$ K. Table \ref{table1}, lists their semimajor axis, radii, surface temperatures (obtained adopting an albedo of 0.01), stellar effective temperatures and stellar radius.   

\begin{figure}
  \begin{center}    
    \subfigure[]{\label{fig2b}\includegraphics[angle=270,width=.45\textwidth]{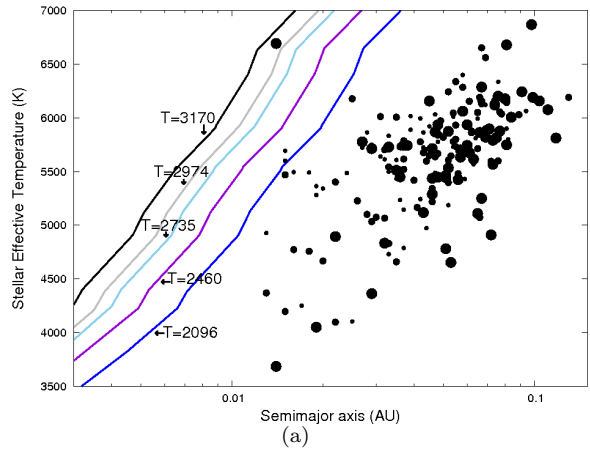}}
    \subfigure[]{\label{fig2a}\includegraphics[angle=270,width=.45\textwidth]{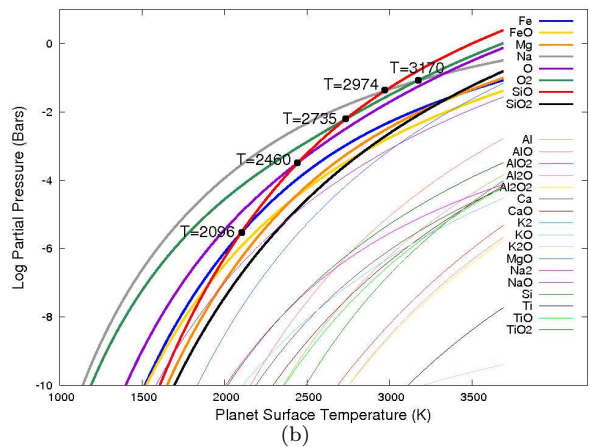}}
  \end{center}
  \caption{$T_{eff,\star}$ vs. $a$ for the Kepler candidates analyzed in this work (upper panel) . KOIs with radius between 2-2.5 R$_{\oplus}$ (large points), radius between 1.5-2 R$_{\oplus}$ (medium points) and $R_p<$1.5 R$_{\oplus}$  (small points). Lines of constant temperatures are shown. Planet surface temperatures  vs. partial pressures of the gases vaporized from a komatiite magma (lower panel). Temperatures at which the dominant gases change are indicated.}
  \label{fig2}
\end{figure}

\begin{deluxetable}{lccccr} 
\tablecolumns{6} 
\tablewidth{0pc} 
\tablecaption{Planetary and stellar parameters for {\it{Kepler}} candidates with radii less than 2.5 R$_{\oplus}$ and temperatures higher than 1000 K.} 
\tablehead{
\colhead{KOI} & \colhead{ $a$ (AU)}   & \colhead{$R_p$ (R$_{\oplus}$)}  & \colhead{ $T_p$ (K)}\tablenotemark{a}  & \colhead{ $T_{eff,\star}$ (K)}    & \colhead{$R_{\star}$ (R$_{\odot}$)}}
\startdata 
   69.01 & 0.056 & 1.6 & 1129 &  5480 &  1.03 \\
   70.02 & 0.046 & 1.6 & 1001 & 5342 & 0.7\\
    72.01 & 0.018 & 1.3 & 1946 & 5491 & 0.98 \\
    85.02 & 0.035 & 1.7 & 1987 & 6006 & 1.66 \\
    85.03 & 0.084 & 2. & 1282 & 6006 & 1.66 \\
    107.01 & 0.075 & 2.1 & 1025 & 5816 & 1.01 \\
    112.02 & 0.048 & 1.7 & 1414 & 5839 & 1.22 \\
    115.02 & 0.076 & 2.2 & 1251 & 6202 & 1.34 \\
    117.02 & 0.058 & 1.3 & 1142 & 5725 & 1 \\
    117.03 & 0.043 & 1.3 & 1326 & 5725 & 1 \\
     123.01 & 0.071 & 2.3 & 1188 & 5897 & 1.25 \\
    124.01 & 0.111 & 2.3 & 1006 & 6076 & 1.32\\
    137.03 & 0.046 & 2.3 & 1335 & 5289 & 1.27 \\
    139.02 & 0.045 & 1.2 & 1272 & 5921 & 0.9 \\
\enddata
\tablecomments{Table \ref{table1} is published in its entirety in the 
electronic edition. A portion is 
shown here for guidance regarding its form and content.}
\tablenotetext{a}{Calculated adopting an albedo of 0.01}
\label{table1}
\end{deluxetable} 
Following \citet{scf09}, we explore the partial pressures of all the gases vaporized at different temperatures, assuming atmospheres free of volatile gases such as H, C, N, S (see discussion). We model the initial planetary atmosphere characteristics depending on the radius and semi-major axis of the Kepler candidates. Figure \ref{fig2b} shows the planet candidates that orbit their star with short periods and accordingly high temperatures. The candidates are shown in different point sizes according to their radius: 2-2.5 R$_{\oplus}$ (large points), 1.5-2 R$_{\oplus}$ (medium points) and candidates with $R_p<$1.5 R$_{\oplus}$ (small points). The lines indicate constant planet surface temperatures. The blue, violet, light blue, grey and black line in figure \ref{fig2b} indicate $T_p$ of 2096, 2460, 2735, 2974 and 3170 K, respectively. These temperatures indicate significant changes in the partial pressures of the dominant gases. Figure \ref{fig2a} shows the partial pressures of the gases as a function of the planet surface temperature. 

Planets located below the blue line in figure \ref{fig2b} have calculated surface temperatures $T_p<$2096 K. As seen in figure \ref{fig2a}, these planets are characterized by partial pressures dominated by Na, O$_2$, O and Fe (from the highest to the lowest partial pressures). In the same way, planets located between the blue and the violet line have temperatures $2096<T_p<2460$ K. In this case, the partial pressure of SiO becomes higher than Fe. With surface temperatures between 2460 and 2735, the planets are located between the violet and the light-blue line. For these surface temperatures, the partial pressure of SiO is higher than O. When the surface temperature is between $2735<T_p<2974$ K (between the light-blue and the grey line), the partial pressure of SiO is higher than O$_2$ and for $2974<T_p<3170$ K (between the grey and the black line) SiO becomes more abundant than monatomic Na. Finally a planet with $T_p> 3170$ K is located above the black line and is characterized by the highest partial pressure of SiO, followed by O$_2$, monatomic O and Na.

\subsubsection{Different types of Atmospheres}\label{section:composiciones}

In order to link the partial pressures to the resulting atmospheric composition, we calculate the column density ($\sigma_i$) of each gas $i$, as a function of the gas' partial pressure, $P_i$, for planetary masses between 1 and 10 M$_{\oplus}$, corresponding to radii between $\sim$ 1-2.5 R$_{\oplus}$ \citep{va07,se07,so07} using equation \ref{sigma}.

\begin{equation}\label{sigma}
\sigma_i=\frac{P_iN_A}{g\mu_i}
\end{equation}

where $N_A$ is Avogadro's number, $\mu_i$ is the molecular weight of the species and $g$ is the gravitational surface acceleration of each planet.  We find that the compositions of the atmospheres do not show significant changes when considering different planetary masses. This means that the surface temperature and therefore the distance to the host star is the dominating factor in determining the composition of the atmosphere, assuming a certain planetary composition.

\begin{figure*}
  \begin{center}
\subfigure[]{\label{fig3b}\includegraphics[angle=270,width=.45\textwidth]{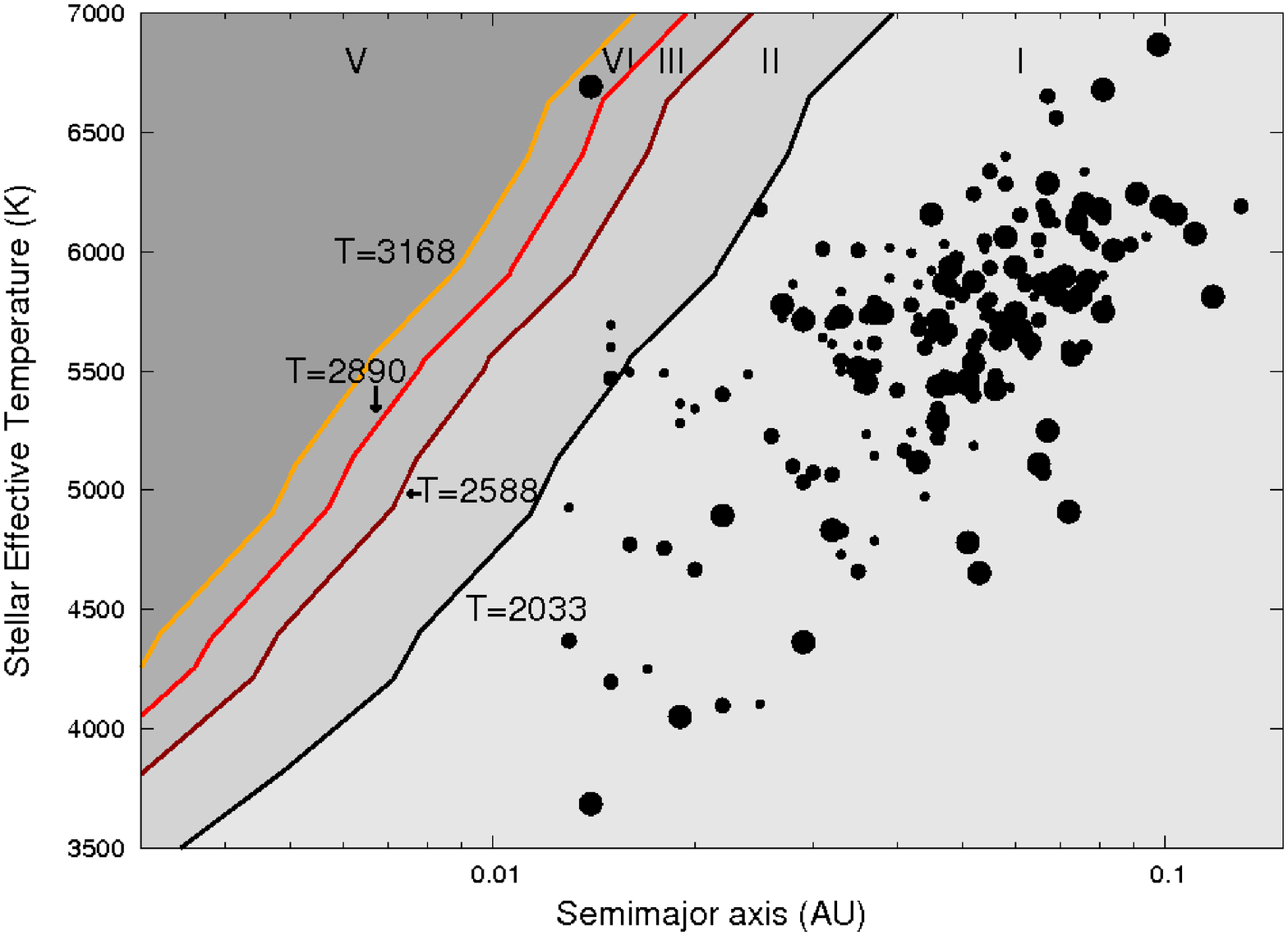}}\subfigure[]{\label{fig3d}\includegraphics[angle=270,width=.45\textwidth]{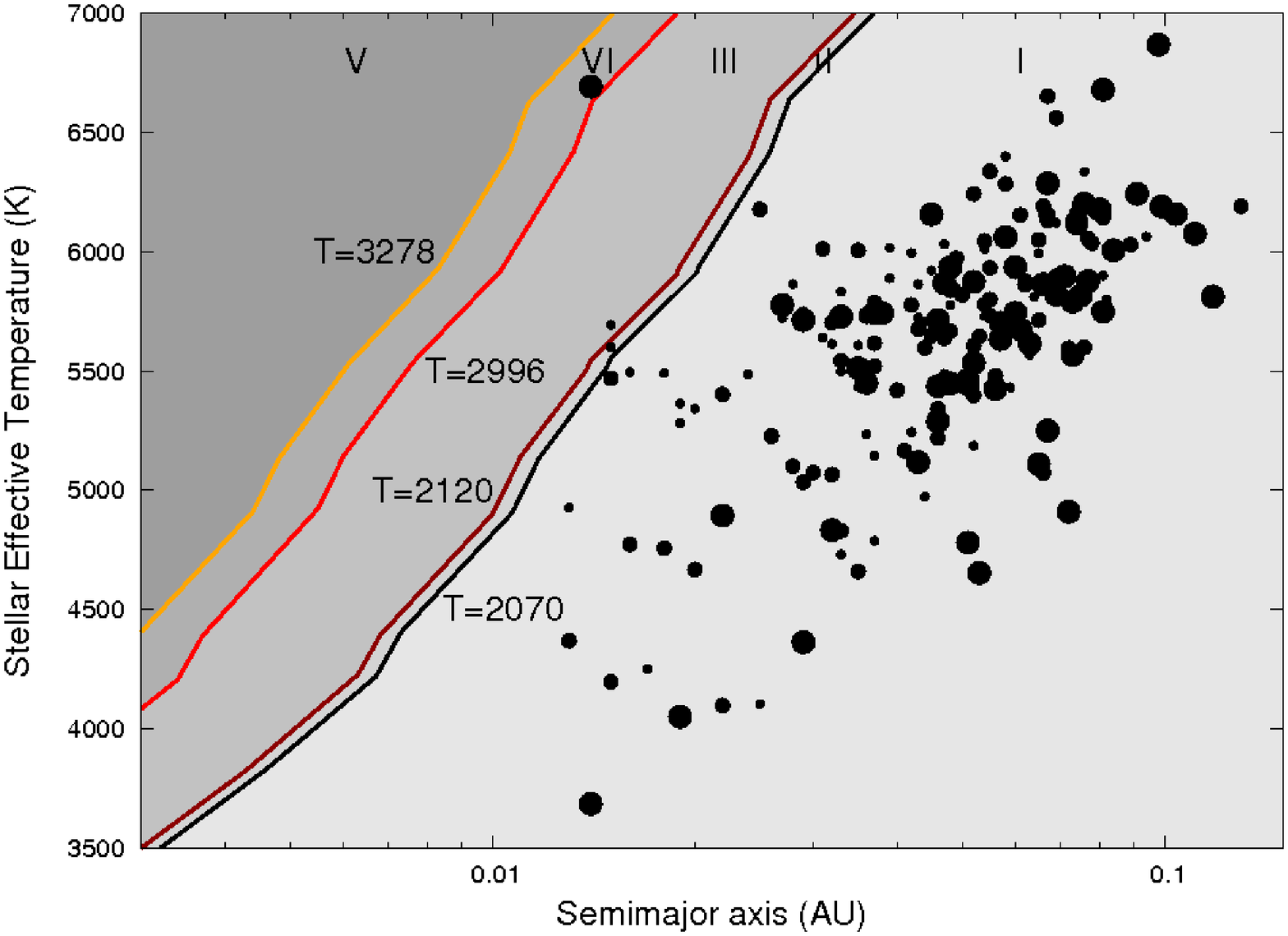}}
\subfigure[]{\label{fig3a}\includegraphics[angle=270,width=.45\textwidth]{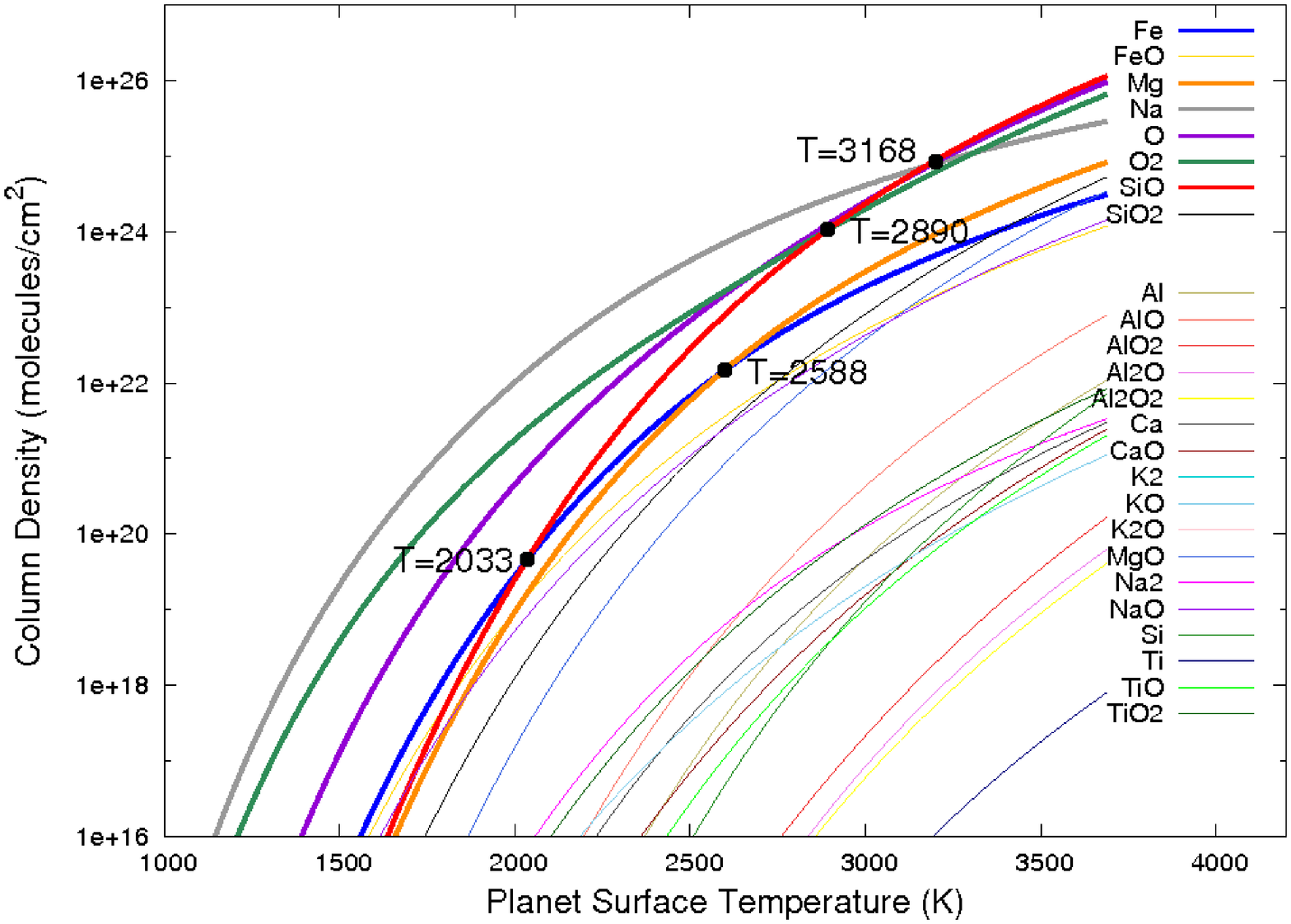}}\subfigure[]{\label{fig3c}\includegraphics[angle=270,width=.45\textwidth]{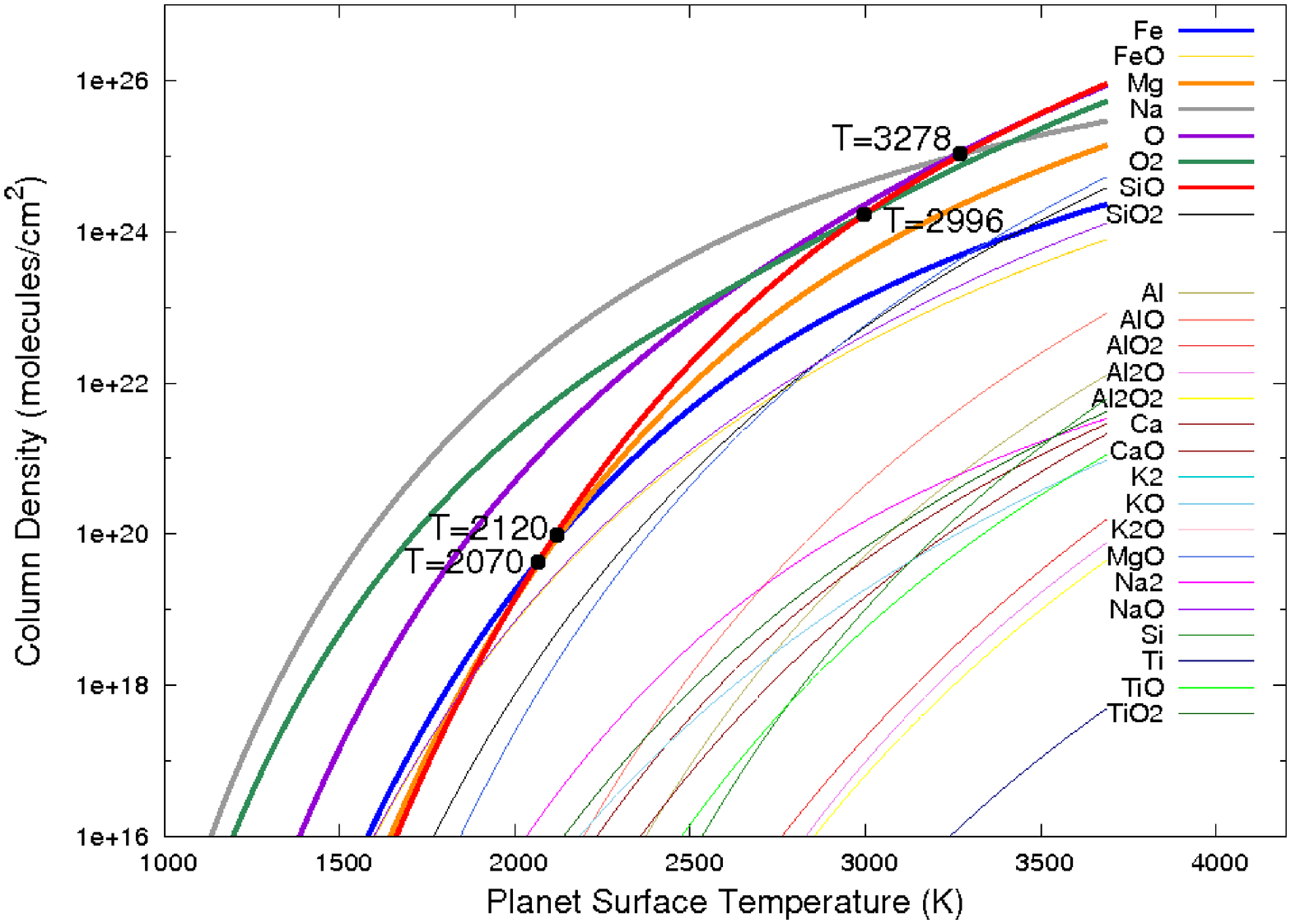}}
 \end{center}
  \caption{Stellar effective temperature vs. semi-major axis of Kepler planetary candidates (upper panels, see figure \ref{fig2b} for detailed explanation) and column densities of the atmospheric gases vs. planet surface temperature (lower panels). The column densities were calculated for a planet of 10 M$_{\oplus}$, but are representative of planets with masses between 1-10 M$_\oplus$  (see section \ref{section:composiciones}). We adopt two different planetary compositions: komatiite (figures \ref{fig3b}, \ref{fig3a}) and bulk silicate of the Earth (figures \ref{fig3d}, \ref{fig3c}).}
  \label{fig3}
\end{figure*}

 We adopt a komatiite composition for hot super-Earths and explore the effects of initial composition on the results by comparing with a Bulk Silicate Earth (B.S.E) initial composition in Figure \ref{fig3}.
Figure \ref{fig3b} (komatiite) and \ref{fig3d} (BSE composition) show the stellar effective temperature vs. the planet's distance to the star.  Lines show the temperatures that separate the different classes of planetary atmospheres. Areas highlighted in gray scale indicate the regions of different types of atmospheres. The planets in each one of these areas are characterized by different types of atmospheres as indicated below. Figures \ref{fig3a} and \ref{fig3c} show the column density of all the gases vaporized vs. the planet surface temperature for a 10 M$_{\oplus}$ planet with komatiite and BSE as the initial composition, respectively.  Since we find similar results for planets with different masses, the column densities shown in Figures \ref{fig3a} and \ref{fig3c} are representative of planets with masses between 1-10 M$_{\oplus}$. The dominant gases are Fe,  Mg, Na, O, O$_2$ and SiO. 

Our results divide hot rocky planet atmospheres into 5 classes. Temperatures that limit each class change when considering different magma compositions. For komatiite composition: 

{\bf{Atmospheres Type I}} ($T_p<2033$ K), characterized by the presence of monatomic Na, O$_2$, monatomic O and monatomic Fe in order of abundance. 

{\bf{Atmospheres Type II}} ($2033<T_p<2588$ K), SiO becomes more abundant than monatomic Fe. The atmosphere is characterized by monatomic Na, O$_2$, monatomic O, SiO, monatomic Fe and monatomic Mg from the most to the least abundant. 

{\bf{Atmospheres Type III}} ($2588<T_p<2890$ K), monatomic Mg becomes more abundant than monatomic Fe. The gases with the highest column densities are monatomic Na, O$_2$, monatomic O, SiO, monatomic Mg and monatomic Fe. Note that the column densities of O$_2$, monatomic O and SiO are almost equal in this temperature range. 

{\bf{Atmospheres Type VI}} ($2890<T_p<3168$ K), SiO becomes more abundant than O$_2$ and the atmosphere becomes silicate and monatomic Na dominated.
 
{\bf{Atmospheres Type V}} ($T_p>3168$ K), dominated by silicate oxide, followed closely by monatomic O, O$_2$ and monatomic Na.

Of the 193 Kepler planetary candidates studied in this work (see Figure \ref{fig3b}), 189 are characterized by Type I atmospheres when adopting komatiite as the initial composition. Three planets are characterized by Type II (between the black and the dark-red line in figure \ref{fig3b}), no planets by Type III (between the dark-red and red lines), only one by Type IV (located between the red and yellow lines) and none by Type V atmospheres (above the yellow line) , if we only consider the irradiation from the star for heating (see discussion). 

When adopting the BSE composition, the temperature limits between the 5 classes change slightly to 2070, 2120, 2996 and 3278 K, respectively. For a BSE composition, 190 planetary candidates are characterized by Type I atmospheres (located below the black line in figure \ref{fig3d}), only 1 planet by Type II atmosphere (between the black and the dark-red lines), 1 by Type III (between the dark-red and red lines), 1 by Type IV  (located between the red and yellow lines) and none by Type V (above the yellow line).  As a result, note that, for this planetary sample, the initial composition is not important when the planet's temperature is high for atmospheres of Type IV and V. Nevertheless, for lower temperatures (Types I, II and III) it becomes an important factor, since a small difference in temperature can change the dominant gases in the planetary atmosphere.

\section{Discussion: the scope of our work}

The composition of the atmosphere depends strongly on the composition of the planetary interior. We show the changes in the characterization when adopting two known compositions taken from our Solar system, but substantially different compositions are possible.

Following \citet{scf09}, we consider atmospheres produced by the outgassing of a volatile-free magma. At these extreme temperatures, elements such as H, C, N and S should escape from the atmosphere in short timescales and consequently are not present in the atmosphere. This assumption is supported by observations of volatile loss in rocky bodies \citep{ss96,fe04} and studies in hot-Jupiters \citep{ba05} and hot super-Earths \citep{le11}. If the volatiles were not lost, the atmosphere would be dominated by volatiles.  In this paper we only concentrate on close-in, hot planetary candidates, consistent with volatile free atmospheres. Cooler planets that could potentially provide habitable conditions are not discussed here (see e.g. \citet{ka11}). 

In this work we have adopted a bond albedo for the planets of $A$=0.01. This assumption is supported by observations \citep{ro09, ro06, ro08} and modeling \citep{su00, bu08, mi10} of hot exoplanets, which find very low albedo. Although there are currently no indications that suggest higher albedos for hot rocky planets, we studied the effect of a higher albedo on our results. When changing the albedo from 0.01 to e.g. 0.2, the size of the sample is affected. For an albedo of 0.2, only 146 planetary candidates with radius less than 2.5 R$_{\oplus}$ have $T_p>1000$ K. Of these 146 planets, 144 are characterized by atmospheres Type I, 1 by Type II, 1 by Type III, 0 by Type IV and 0 by Type V (adopting komatiite composition). 
 
We assume here that the planets are only heated by the irradiation of the star. Other heat sources like short-lived radioisotopes, impacts, core formation and tidal effects can increase the surface temperature. \citet{he09} showed that
the insolation is the most important heating source for planets with periods larger than $\sim$ 2 days and for planets with shorter periods and non eccentric orbits. For very eccentric planets ($e>$0.1) which orbit their star with periods less than $\sim$ 2 days, tidal heating dominates over insolation. We explore  the effects of tidal heating in the Kepler planetary candidates with $P<$ 2 days, using eq. \ref{tides} \citep{he09,md05},

\begin{equation}\label{tides}
T_{p,tidal+ins}=\bigg(\frac{\pi R_p^2(1-A)\frac{L_{\star}}{4\pi a^2}+\frac{21}{2}\frac{k_2}{Q}\frac{G R_p^5 n e^2}{a^6}M_{\star}^2}{4 \pi R_p^2 \sigma}\bigg)^{\frac{1}{4}}
\end{equation}
where, $k_2$ is the second-order Love number of the planet. For a super-Earth we adopt 0.3 (Henning personal communication), $Q$ is the quality factor of the planet, $n$ its mean motion and $e$ the eccentricity, which is unknown for the Kepler candidates. We adopt different values for $e$ and $Q$. Most of the candidates that might be affected are located very close to the star ($a$ less than $\sim$ 0.03 AU), where we expect almost circular orbits. We adopt $e=$0.001 and $e=$0.01 and two different values for $Q$: 200 and 20000, which represent planets with limited partial melting or significant partial melting in the mantle, respectively  \citep{he09}, to explore this effect. In the four cases analyzed, we find that there is no significant change in the surface temperatures of these planets when considering heating by insolation plus tidal effects compared to heating by insolation only.  The role of tides in these planets, if occurring, will be to increase the convective vigor of the mantle, causing a larger fraction of total mantle volume to be accessible to surface degassing, but will not change the surface temperature significantly.

\section{Conclusions}

In this work we explore the different atmospheric composition of hot super-Earth candidates from the February {\it{Kepler}} database according to their radii, semi-major axis and stellar effective temperature. 615 Kepler planet candidates have radii less than 2.5 R$_{\oplus}$, of which 193 have calculated surface temperatures larger than 1000 K (assuming an albedo of 0.01), and are the focus of this work. We model the initial planetary atmospheric compositions for komatiite and bulk silicate Earth magma compositions. The planet surface temperatures were calculated assuming insolation only. We discuss the consequences of adopting different sources of heating and find that tidal heating is not important when adopting low eccentricities, consistent with the {\it{Kepler}} sample. 

The results indicate 5 types of atmospheres, based on the abundance of the dominant gases (Fe, Mg, Na, O, O$_2$ and SiO). When adopting komatiite composition, 189 of 193 {\it{Kepler}} candidates present a Type I model atmosphere dominated by monatomic Na, O$_2$, monatomic O and monatomic Fe. Three planetary candidates are characterized by Type II atmospheres, where SiO becomes more abundant than monatomic Fe, no candidates with Type III atmospheres, where monatomic Mg is more abundant than monatomic Fe, and only one planet is characterized by Type IV, with a large abundance of SiO.  Finally, there is no Kepler planetary candidate with an atmosphere of Type V, which is completely dominated by SiO.  

This atmospheric characterization is almost independent of the planetary mass between 1-10 M$_{\oplus}$ and low eccentricities. The initial magma composition is important for the resulting atmospheric gases, especially for lower surface temperatures. We find that the distance from the host star is the dominant factor to characterize a rocky planet's atmosphere.
  
\acknowledgments{
We are grateful to Dimitar Sasselov and Wade Henning for stimulating discussions. }

\end{document}